\documentclass[preprint,aps]{revtex4}
\newcommand{\be}{\begin{equation}}
\newcommand{\ee}{\end{equation}}
\newcommand{\ba}{\begin{eqnarray}}
\newcommand{\ea}{\end{eqnarray}}

\begin{document}

\draft

\title{Dark Matter Haloes and Rotation Curves \\
via Brans-Dicke Theory} 
       
\author{Hongsu Kim\footnote{e-mail : hongsu@astro.snu.ac.kr}}

\address{Astronomy Program, SEES, Seoul National University, Seoul, 151-742, KOREA}


\begin{abstract}
In the present work, the Brans-Dicke (BD) theory of gravity is taken as a possible theory of k-essence. 
Then starting with the (already known) Brans-Dicke-Schwarzschild solution which can
represent the gravitationally bound static configurations of the BD scalar k-essence, issues like
whether these configurations can reproduce the observed properties of galactic dark matter haloes
have been addressed. It has been realized that indeed the BD scalar k-essence can cluster into
dark matter halo-like objects with flattened rotation curves while exhibiting a dark energy-like
negative pressure on larger scales.

\end{abstract}


\maketitle

\narrowtext

\newpage

\begin{center}
{\rm\bf I. Introduction}
\end{center}

Currently, perhaps the most fashionable candidates for the unified model of 
dark matter and dark energy \cite{wmap} with non-trivial dynamics are quintessence and k-essence. 
The main difference
between the two models is that the quintessence models \cite{quintessence} involve canonical kinetic terms
and the sound speed of $c^2_{s} = 1$ while the k-essence models \cite{kessence} employ rather exotic scalar fields
with non-canonical (non-linear) kinetic terms which typically lead to the {\it negative} pressure.
And the most remarkable property of these k-essence models is that the typical k-essence field can
overtake the matter energy density and induce cosmic acceleration only at the onset of the 
matter-dominated era and particularly at about the present epoch. These models are also expected to provide
a successful explanation of the phenomena associated with the dark matter. In the present work, we take
the Brans-Dicke (BD) theory of gravity \cite{bd} as a possible k-essence theory since it involves probably 
the simplest form of such non-linear kinetic term for the (BD) scalar field.
Besides, the BD scalar field (and the BD theory itself) is not of quantum origin. Rather it is
classical in nature and hence can be expected to serve as a very relevant candidate to play some role in 
the late-time evolution of the universe such as the present epoch.
Indeed, the BD theory is the most studied and hence the best-known of all the alternative theories of
classical gravity to Einstein's general relativity \cite{will}. 
This theory can be thought of as a minimal extension of general relativity designed to properly 
accomodate both Mach's principle \cite{will, weinberg} and Dirac's 
large number hypothesis \cite{will, weinberg}. Namely, the theory employs the viewpoint in which the Newton's
constant $G$ is allowed to vary with space and time and can be written in terms of a scalar (``BD scalar'')
field as $G = 1/\Phi$. As a scalar-tensor theory of gravity, it involves an adjustable but undetermined
``BD-parameter'' $\omega$ and as is well-known, the larger the value of $\omega$, the more dominant the
tensor (curvature) degree and the smaller the value of $\omega$, the larger the effect of the BD scalar.
And as long as we select sufficiently large value of $\omega$, the predictions of the theory agree
perfectly with all the observations/experiments to date \cite{will}. For this reason, the BD theory has
remained a viable theory of classical gravity. 
However, no particularly overriding reason thus far has ever emerged
to take it seriously over the general relativity. As shall be presented shortly in this work, here
we emphasize that it is the existence of dark matter (and dark energy as well, see \cite{hongsu2}) that
puts the BD theory over the general relativity as a more relevant theory of classical gravity consistent
with observations that have so far been unexplained within the context of general relativity. \\

\begin{center}
{\rm\bf II. Haloes of BD scalar k-essence}
\end{center}

In general, the Brans-Dicke theory of gravity is described, in the absence of ordinary matter, by the action 
\begin{eqnarray}
S = \int d^4x \sqrt{g}{1\over 16\pi}\left[\Phi R - \omega {{\nabla_{\alpha}\Phi
\nabla^{\alpha}\Phi }\over \Phi}\right]
\end{eqnarray}
where $\Phi $ is the BD scalar field representing the inverse of Newton's constant which is allowed to
vary with space and time and $\omega $ is the generic dimensionless parameter of the
theory. Extremizing this action then with respect to the metric $g_{\mu \nu}$ and the
BD scalar field $\Phi $ yields the classical field equations given respectively by 
\begin{eqnarray}
G_{\mu \nu} &=& R_{\mu \nu} - {1\over 2}g_{\mu \nu}R = 8\pi T^{BD}_{\mu \nu}, 
~~~\nabla_{\alpha}\nabla^{\alpha}\Phi = 0  ~~~{\rm where}  \nonumber \\
T^{BD}_{\mu \nu} &=& {1\over 8\pi}\left[{\omega \over \Phi^2}(\nabla_{\mu}\Phi \nabla_{\nu}\Phi
- {1\over 2}g_{\mu \nu}\nabla_{\alpha}\Phi \nabla^{\alpha}\Phi) + {1\over \Phi}(\nabla_{\mu}
\nabla_{\nu}\Phi - g_{\mu \nu}\nabla_{\alpha}\nabla^{\alpha}\Phi)\right].  
\end{eqnarray}
Note that here in the present work, we are interested in the role played by the BD scalar
field (i.e., a k-essence) as a dark matter particularly in forming galactic dark matter haloes
inside of which the well-known rotation curves have been observed.
Since the galactic dark matter haloes are {\it roughly} static and spherically-symmetric, we first
should look for such dark matter halo-like solution of these BD field equations.
Interestingly enough, the Brans-Dicke-Schwarzschild (BDS) spacetime solution to these vacuum BD field
equations that happens to meet our above-mentioned needs has been found some time ago \cite{hongsu1} in 
rather a theoretical attempt to construct non-trivial black hole spacetime solutions in BD theory.
To summarize, the BDS spacetime solution that can be obtained by setting
$a = e = 0$ in the Brans-Dicke-Kerr-Newman (BDKN) solution in eq.(11) of \cite{hongsu1} takes the form
\begin{eqnarray}
ds^2 &=& \Delta^{-2/(2\omega+3)}\sin^{-4/(2\omega+3)}\theta
\left[-\left(1 - {2M\over r}\right)dt^2 + r^2 \sin^2 \theta d\phi^2 \right] \nonumber \\
&+& \Delta^{2/(2\omega+3)}\sin^{4/(2\omega+3)}\theta
\left[\left(1 - {2M\over r}\right)^{-1}dr^2 + r^2 d\theta^2 \right], \\
&\Phi& (r, \theta) = \Delta^{2/(2\omega+3)}\sin^{4/(2\omega+3)}\theta \nonumber
\end{eqnarray}
where $\Delta = r(r - 2M)$. A remarkable feature of this BDS solution is the fact that,
unlike the Schwarzschild solution in general relativity, the spacetime it describes is 
static (i.e., non-rotating) but {\it not} spherically-symmetric. Of great interest in this earlier
construction was the realization that non-trivial black hole solutions {\it different} from
general relativistic solutions could occur in this BD theory for the generic BD-parameter values
in the range $-5/2 \leq \omega <-3/2$ \cite{hongsu1}. In the present study, however, since we are
interested in the galactic halo-like configuration, we do {\it not} want this BDS solution to become 
``black'' and this amounts to considering the BDS solution having the value of $\omega$-parameter
well outside this range. Besides, we are only interested in whether
the self-gravitating k-essence, i.e., the BD scalar field can generally cluster into dark matter 
halo-like objects which would be the gravitationally bound static solution configurations of 
super-galactic scale (i.e., the large but finite-$r$ behavior). 
Therefore, the peculiar microscopic geometrical nature of this 
BDS solution such as the issue of regularity of the potential Killing horizon (i.e., the finiteness of
the invariant curvature polynomials there) addressed in \cite{hongsu1} or that of seemingly failure
of asymptotic flatness and internal infinity nature
of the symmetry axis discussed in \cite{hongsu3} are all irrelevant for the present purposes.  \\
Therefore first, it appears that the BD scalar k-essence can indeed cluster into halo-like 
configurations as it can be represented by the BDS solution. Our natural next mission is then to ask 
whether these configurations really can reproduce the properties of dark matter haloes, namely if our 
BD scalar k-essence model for dark matter can reproduce the flattening of the rotation velocity curves
inside these halo configurations consistent with the observations. Thus we now attempt to obtain the
rotation curves in our BD scalar k-essence halo.
Since henceforth we need concrete ``numbers'', we now restore both Newton's constant $G_{0}$ and the
speed of light $c$ in order to come from the geometrical unit ($G_{0} = c = 1$) back to the CGS (or MKS)
unit. Then the energy-momentum tensor of the BD scalar field given earlier in eq.(2) now takes the
form
\begin{eqnarray}
T^{BD}_{\mu \nu} &=& {c^4\over 8\pi G_{0}}\left[{\omega \over \Phi^2}(\nabla_{\mu}\Phi \nabla_{\nu}\Phi
- {1\over 2}g_{\mu \nu}\nabla_{\alpha}\Phi \nabla^{\alpha}\Phi) + {1\over \Phi}(\nabla_{\mu}
\nabla_{\nu}\Phi - g_{\mu \nu}\nabla_{\alpha}\nabla^{\alpha}\Phi)\right]
\end{eqnarray}
and in the BDS solution in eq.(3) above, we should replace $M\rightarrow G_{0}M/c^2 \equiv \tilde{M}$
where $G_{0}$ denotes the present value of the Newton's constant.
Apparently, (4) has the dimension of the energy-momentum density in the CGS unit, $(erg/cm^3)$. 
We now turn to the computation of energy
density profile and (anisotropic) pressure components of the k-essence playing the role of the dark 
matter by treating the BD scalar field as a (dark matter) {\it fluid}. 
The BD scalar field fluid, however, would
fail to be a ``perfect'' fluid as can readily be envisaged from the fact that the associated BDS solution
configuration is not spherically-symmetric. Namely, its pressure cannot be ``isotropic'', i.e.,
$P_{r} \neq P_{\theta} \neq P_{\phi}$. Such fluid may be called {\it imperfect} fluid due to the
{\it anisotropic} pressure components and as such its stress tensor can be written as
\begin{eqnarray}
T^{BD ~\mu}_{\nu} = 
\pmatrix{-c{^2}\rho & 0 & 0 & 0 \cr
         0 & P_{r} & T^{r}_{\theta} & 0 \cr
         0 & T^{\theta}_{r} & P_{\theta} & 0 \cr
         0 & 0 & 0 & P_{\phi} \cr}.
\end{eqnarray}
And it is to be contrasted to its counterpart of the usual perfect fluid with isotropic pressure 
given by the well-known form,
$T^{\mu}_{\nu} = P\delta^{\mu}_{\nu} + (c^2 \rho + P)U^{\mu}U_{\nu} =
diag (-c^2 \rho, ~P, ~P, ~P)$ where $U^{\alpha} = dX^{\alpha}/d\tau $ (with $\tau $ being the proper time)
denotes the 4-velocity of the fluid element normalized such that $U^{\alpha}U_{\alpha} = -1$.
Note that in addition to the diagonal entries representing the (anisotropic) pressure components
$T^{i}_{i}=P_{i}$ (where no sum over $i$), there are off-diagonal entries $T^{r}_{\theta}$,
$T^{\theta}_{r}$ representing a {\it shear stress} which also results from the failure of
spherical symmetry.
Thus by substituting the BDS solution given in eq.(3) into the BD
energy-momentumm tensor in eq.(4) and then setting (4) equal to (5), we can eventually read off
the energy density and the pressure components of the BD scalar field imperfect fluid to be
\begin{eqnarray}
\rho &=& {c^2\over 8\pi G_{0}}{4 \over (2\omega + 3)^2}{1\over r^2 \Delta}
\Delta^{-2/(2\omega+3)}\sin^{-4/(2\omega+3)}\theta  \nonumber \\
&\times & \left[2(\omega +1)\left\{(r-\tilde{M})^2 + \Delta \cot^2 \theta \right\} 
- (2\omega +3)\tilde{M}(r-\tilde{M})\right], \nonumber \\
P_{r} &=& -{c^4\over 8\pi G_{0}}{4 \over (2\omega + 3)^2}{1\over r^2 \Delta}
\Delta^{-2/(2\omega+3)}\sin^{-4/(2\omega+3)}\theta  \nonumber \\
&\times& \left[2(\omega +2)(r-\tilde{M})^2 + 2(\omega -1)\Delta \cot^2 \theta 
- (2\omega +3)\left\{\Delta + \tilde{M}(r-\tilde{M})\right\} \right], \nonumber \\
P_{\theta} &=& {c^4\over 8\pi G_{0}}{4 \over (2\omega + 3)^2}{1\over r^2 \Delta}
\Delta^{-2/(2\omega+3)}\sin^{-4/(2\omega+3)}\theta 
\left[2(\omega -1)\left\{\Delta \cot^2 \theta - (r-\tilde{M})^2\right\}\right. \nonumber \\ 
&+& \left. (2\omega +3)(r-\tilde{M})(r-2\tilde{M})
+  \left\{4\cos^2 \theta - (2\omega +3)\right\}{\Delta \over \sin^2 \theta } 
\right], \\
P_{\phi} &=& -{c^4\over 8\pi G_{0}}{4 \over (2\omega + 3)^2}{1\over r^2 \Delta}
\Delta^{-2/(2\omega+3)}\sin^{-4/(2\omega+3)}\theta \nonumber \\
&\times& \left[2(\omega +1)(r-\tilde{M})^2 - \Delta \cot^2 \theta  
- (2\omega +3)(r-\tilde{M})(r-2\tilde{M})\right] \nonumber \\
T^{r}_{\theta} &=& \Delta T^{\theta}_{r} = 
{c^4\over 8\pi G_{0}}{4 \over (2\omega + 3)^2}{1\over r^2 }\cot \theta
\Delta^{-2/(2\omega+3)}\sin^{-4/(2\omega+3)}\theta \nonumber \\
&\times& \left[4\omega (r-\tilde{M}) - (2\omega +3)(r-2\tilde{M})\right]. \nonumber 
\end{eqnarray}
Note that the off-diagonal components $T^{r}_{\theta}$, $T^{\theta}_{r}$ are {\it odd} functions
of $\theta $ while the diagonal components $(\rho, P_{r}, P_{\theta}, P_{\phi})$ are {\it even}
functions of the polar angle under $\theta \rightarrow (\pi - \theta )$. As a result, the off-diagonal
components vanish (i.e., no shear stress survives) if we average over this polar angle to get a net
stress.
Thus, first the equation of state of this BD scalar k-essence fluid forming a galactic halo is given by
\begin{eqnarray}
w = {P\over c^2\rho } 
= -\frac{\left[2(\omega +2)(r-\tilde{M})^2 + 2(\omega -1)\Delta \cot^2 \theta 
- (2\omega +3)\left\{\Delta + \tilde{M}(r-\tilde{M})\right\}\right]}
{2(\omega +1)\left\{(r-\tilde{M})^2 + \Delta \cot^2 \theta \right\} 
- (2\omega +3)\tilde{M}(r-\tilde{M})} 
\end{eqnarray}
where $P = P_{r}$. Namely, $P = w(r, \theta)c^2 \rho $ with
$w(r, \theta) \sim O(1)$ meaning that this k-essence fluid is essentially a {\it barotropic}
fluid but with ``position-dependent'' coefficient $w(r, \theta)$. Note that although the BD scalar
k-essence is a candidate for dark matter, it is not quite a dust.
In principle, the speed of sound in this BD scalar field fluid can also be evaluated via
$c^2_{s} = dP/d\rho$  
but we shall not discuss it in any more detail in this work.
We are now ready to compute the behavior of rotation curves in the outer region 
(i.e., at large but finite-$r$, say, $r >> G_{0}M/c^2$ ) of our BD scalar k-essence halo. 
To be more precise, for a galaxy of typical (total) mass $M\sim 10^{11}M_{\odot}$, the outer region of its
dark matter halo, say, $r\sim 10 (kpc) \simeq 10^{23}(cm)$ is much greater than 
$G_{0}M/c^2 \simeq 10^{16}(cm)$ by a factor of $10^{7}$.
Thus to this end, we first approximate the expressions for the energy density and the (radial) 
pressure of the k-essence given in eq.(6) for large-$r$. They are
\begin{eqnarray}
\rho &\simeq & {c^2\over 8\pi G_{0}}{8(\omega + 1) \over (2\omega + 3)^2}{1\over r^2 \sin^2 \theta }
\Delta^{-2/(2\omega+3)}\sin^{-4/(2\omega+3)}\theta ,   \\
P &\simeq & -{c^4\over 2\pi G_{0}}{1 \over (2\omega + 3)^2}{1\over r^2}\left[2(\omega -1)\cot^2 \theta + 1\right]
\Delta^{-2/(2\omega+3)}\sin^{-4/(2\omega+3)}\theta . \nonumber 
\end{eqnarray}
Note that in the above approximations and in the discussions below, it was and it shall be
assumed that the metric function $\Delta = r(r - 2\tilde{M}) \simeq r^2$ for large-$r$.
It is interesting to note that as a ``k-essence'' constituting a dark matter halo, the energy
density $\rho $ of the BD scalar field is almost certainly {\it positive everywhere} (i.e., for
both small and large-$r$). In the mean time, its (radial) pressure $P$ particularly at larger scale 
(i.e., for large-$r$) turns out to be {\it negative} although its sign appears unclear at small
scale (i.e., for small-$r$). \\
Finally, we are ready to determine the rotation curve inside our BD scalar k-essence halo.
First in the most naive sense, the apparent rotation velocity of an object at radius $r$ from the
galactic center is given by the Kepler's third law, $v^2 = G_{0}M(r)/r$. Thus for our case, using
the BD scalar k-essence energy density profile given earlier, we have $M(r) = \int^{2\pi}_{0}d\phi 
\int^{\pi}_{\epsilon}d\theta \int^{r}_{0}dr \sqrt{g_{rr}g_{\theta \theta}g_{\phi \phi}}\rho (r, \theta) =
\left(2c^2/G_{0}\right)\left[(\omega +1)/(2\omega +1)(2\omega +3)\right]f(\omega)r r^{-2/(2\omega +3)}$ 
and hence
\begin{eqnarray}
v^2(r) = \frac{G_{0}M(r)}{r} = c^2 \frac{2(\omega +1)}{(2\omega +1)(2\omega +3)}f(\omega)
r^{-2\over (2\omega +3)} 
\end{eqnarray}
where $f(\omega) \equiv \int^{\pi}_{\epsilon}d\theta \sin^{-[1+2/(2\omega +3)]}\theta
= 2\int^{1-\delta}_{0}dx [1-x^2]^{-(2\omega +4)/(2\omega +3)}$ with $\epsilon, ~\delta <<1$. 
(Note here that the integration over the polar angle $\theta $ starts not from $0$ but from $\epsilon <<1$
as the symmetry axis $\theta = 0$ of the BDS solution in eq.(3) possesses an internal infinity nature,
namely, the symmetry axis is infinite proper distance away as discussed carefully in \cite{hongsu3}.)
It has been known for some time that in order for the BD theory to remain a viable theory of classical
gravity passing all the observational/experimental tests to date, the BD $\omega$-parameter has to have
a large value, say, $|\omega |\geq 500$ \cite{will}. 
In our previous study \cite{hongsu1}, in the mean time, it has
been realized that the static solution to the vacuum BD field equations
given in eq.(3) above can turn into a black hole spacetime for $-5/2 \leq \omega <-3/2$. Thus now for
$|\omega |\geq 500$, the same static solution eq.(3) we are considering represents just a halo-like
configuration with regular geometry everywhere (i.e., having no horizon) which is static but not
exactly spherically-symmetric (note that the galactic haloes are also believed to be nearly
spherically-symmetric but not exactly).
Thus if we substitute a large-$\omega$ value, say, $\omega \sim 10^6$ into eq.(9) above, evidently
$M(r) \sim r$ and hence we get
\begin{eqnarray}
v(r) \simeq 100 (km/s) \times r^{-(1/10^6)}
\end{eqnarray}
since for $\omega \sim 10^6$, $f(\omega) \simeq O(1)$.
Namely for this large-$\omega$ value, the rotation curve gets flattened out as
$r^{-(10^{-6})} \sim constant $ and its magnitude becomes several hundred
$(km/s)$. Indeed, this is in impressive agreement with the data for rotation curves
observed in spiral/elliptic galaxies with $M/L \simeq (10 - 20) M_{\odot}/L_{\odot}$ and
in low-surface-brightness (LSB)/dwarf galaxies with $M/L \simeq (200 - 600) M_{\odot}/L_{\odot}$
(where $M/L$ denotes the so-called ``mass-to-light'' ratio given in the unit of solar mass-to-luminosity
ratio exhibiting the large excess of dark matter over the luminous matter) \cite{rc}. 
Rotation curves are observed usually via the
measurements of the Doppler shift of the $21 cm$ emission line from neutral hydrogen (HI)
for distant galaxies and of the light emitted by stars for nearby galaxies \cite{halo1, halo2}. 
It is also interesting to note that this behavior of the rotation curve in our BD theory
k-essence dark matter halo model is {\it independent} of the mass of the host galaxy as it should be. 
Namely, this behavior of the rotation curve comes exclusively from the nature
of the dark matter, i.e., the BD scalar field k-essence. 
We also point out that even if we employ more careful expression for the rotation velocity curve involving
the Doppler shift of light emitted by the orbiting objects (assuming that the k-essence halo is almost
spherically-symmetric), namely $v^2(r) = G_{0}M(r)/r + 4\pi r^2 G_{0}P/c^2$ \cite{khalo} 
(with $P$ being the radial pressure given in eq.(8) above), the conclusions above remain the same. \\
Next, the equation of state in eq.(7) of this BD scalar k-essence becomes, 
in the outer region of the galactic dark matter halo (i.e., at large-$r$),
\begin{eqnarray}
w \simeq - \frac{\left[2(\omega -1)\cos^2 \theta + \sin^2 \theta \right]}
{2(\omega + 1)}
\end{eqnarray}
which is obviously negative due to the {\it negative} pressure (and still {\it positive}
energy density) in this outer region. Moreover, for the large-$\omega$ value, 
i.e., $\omega \sim 10^6$ for which the rotation curve gets flattened out that we just have
realized, this equation of state at large-$r$ further approaches
$w \simeq - \cos^2 \theta \simeq - O(1)$. (Incidentally, it is interesting to note that in the vicinity
of the equatorial plane $\theta = \pi/2$, $w = 0$, namely, the BD scalar k-essence
behaves like nearly a dust.)
This observation is particularly interesting as
it appears to indicate that the BD scalar k-essence we are considering possesses
{\it dark energy-like} negative pressure on larger scales. And this observation is indeed 
consistent with our previous study \cite{hongsu2} that on the cosmological scale, the BD 
scalar field does exhibit the nature of dark energy possessing the negative pressure. 

\begin{center}
{\rm\bf III. Concluding remarks}
\end{center}

In the present work, starting with the (already known) BDS solution which can
represent the gravitationally bound static configurations of the BD scalar k-essence, issues like
whether these configurations can reproduce the observed properties of galactic dark matter haloes
have been investigated. It has been realized that indeed the BD scalar k-essence can cluster into
dark matter halo-like objects with flattened rotation curves while exhibiting a dark energy-like
negative pressure on larger scales.
Thus to conclude, from this success of ``BD scalar field as a k-essence'' to account 
for the asymptotic flattening of galaxy rotation curves while forming galactic 
dark matter haloes {\it plus} the original spirit of the BD theory in which the BD scalar field is 
prescribed {\it not} to have direct interaction with ordinary matter fields (in order not to interfere 
with the great success of equivalence principle), we suggest that the Brans-Dicke theory of gravity 
{\it is} a very promising theory of dark matter. And this implies, among others, that dark matter
(and dark energy as well, see \cite{hongsu2}) might not be some kind of unknown exotic ``matter'', but
the effect resulting from the space-time varying nature of the Newton's constant represented by a
(k-essence) scalar field. Even further, this successful account of the
phenomena associated with the dark matter of the present universe via the BD gravity theory might be 
an indication that the truly relevant theory of classical gravity at the present epoch is not
general relativity but its simplest extension, the Brans-Dicke theory with its generic parameter value
$\omega \sim 10^6$ fixed by the dark matter observation ! \\

This work was financially supported by the BK21 Project of the Korean Government.


\noindent

\begin{center}
{\rm\bf References}
\end{center}


\begin{thebibliography}{99}

\bibitem{wmap} D. N. Spergel et al., Astrophys. J. Suppl. {\bf 148}, 175 (2003).

\bibitem{quintessence} B. Ratra and P. J. E. Peebles, Phys. Rev. {\bf D37}, 3406 (1988) ;
                       J. Frieman, C. Hill, A. Stebbins, and I. Waga, Phys. Rev. Lett. {\bf 75}, 2077 (1995) ;
                       R. R. Caldwell, R. Dave, and P. J. Steinhardt, Phys. Rev. Lett. {\bf 80}, 1582 (1998).

\bibitem{kessence} C. Armendariz-Picon, V. Mukhanov, and P. J. Steinhardt, Phys. Rev. Lett.
                   {\bf 85}, 4438 (2000) ; Phys. Rev. {\bf D63}, 103510 (2001) ;
                   M. Malquarti, E. J. Copeland, A. R. Liddle, and M. Trodden, Phys. Rev. {\bf D67}, 123503 (2003).
                   
\bibitem{bd} C. Brans and C. H. Dicke, Phys. Rev. {\bf 124}, 925 (1961).

\bibitem{will} C. M. Will, {\it Was Einstein Right ?}, 
                   (Basic Books, Inc., Publishers/New York, 1986). 

\bibitem{weinberg} S. Weinberg, {\it Gravitation and Cosmology}, Ch.16, p 619-631, 
                   (John Wiley and Sons, Inc., 1972). 

\bibitem{hongsu2} H. Kim, Phys. Lett. {\bf B606}, 223 (2005).

\bibitem{hongsu1} H. Kim, Phys. Rev. {\bf D60}, 024001 (1999).

\bibitem{hongsu3} H. Kim and H. M. Lee, gr-qc/0402094, Int. J. Mod. Phys. A, to appear.

\bibitem{rc} See for instance, over 100 rotation curve fits given in
             J. R. Brownstein and J. W. Moffat, astro-ph/0506370 ; see also, V. Sahni, astro-ph/0403324.

\bibitem{halo1} S. McGaugh, V. Rubin, and E. de Block, Astron. J. {\bf 122}, 2381 (2001).
                  
\bibitem{halo2} M. Persic, P. Salucci, and F. Stel, Mon. Not. Roy. Astron. Soc. {\bf 281}, 27 (1996) ;
                E. Corbelli and P. Salucci, astro-ph/9909252 ;
                Y. Sofue and V. Rubin, astro-ph/0010594.

\bibitem{khalo} C. Armendariz-Picon and E. A. Lim, astro-ph/0505207.

\end{thebibliography}
\end{document}